\documentclass[a4paper,fleqn,usenatbib]{mnras}

\usepackage{newtxtext,newtxmath}

\usepackage[T1]{fontenc}
\usepackage{ae,aecompl}

\usepackage{graphicx}	
\usepackage{amsmath}	
\usepackage{amssymb}	

\title[V1500 Cygni]{Asynchronous polar V1500 Cyg: orbital, spin and beat periods}

\author[E. P. Pavlenko et al.]{
E. P. Pavlenko,$^{1,2}$\thanks{E-mail: eppavlenko@gmail.com}
P. A. Mason,$^{3,4}$
A. A. Sosnovskij,$^{1}$
S. Yu. Shugarov,$^{5,6}$
\newauthor
Ju. V. Babina,$^{1}$
K. A. Antonyuk,$^{1}$
M. V. Andreev,$^{7,8}$
N. V. Pit,$^{1}$,
O. I. Antonyuk,$^{1}$ 
\newauthor
A. V. Baklanov$^{1}$
\\
$^{1}$Federal State Budget Scientific Institution Crimean Astrophysical Observatory of RAS, Nauchny, 298409, Republic of Crimea, Russia\\
$^{2}$V. I. Vernadsky Crimean Federal University, 4 Vernadskogo Prospekt, Simferopol, 295007, Republic of Crimea, Russia\\
$^{3}$New Mexico State University, MSC 3DA, Las Cruces, NM, 88003, USA\\
$^{4}$Picture Rocks Observatory, 1025 S. Solano, Suite D, Las Cruces, NM, 88001, USA\\$^{5}$Astronomical Institute of Slovak Academy of Sciences,  05960 Tatranska Lomnica, Slovakia\\
$^{6}$Sternberg Astronomical Institute, Lomonosov Moscow State University, Universitetsky Ave., 13, Moscow 119992, Russia\\
$^{7}$Terskol Branch of Institute of Astronomy, Russian Academy of Sciences, 361605, Peak Terskol, Kabardino-Balkaria Republic, Russia\\
$^{8}$International Center for Astronomical, Medical and Ecological Research of NASU, Ukraine 27 Akademika Zabolotnoho Str. 03680 Kyiv, Ukraine}

\date{Accepted XXX. Received YYY; in original form ZZZ}

\pubyear{2017}

\begin{document}
\label{firstpage}
\pagerange{\pageref{firstpage}--\pageref{lastpage}}
\maketitle

\begin{abstract}
	
The bright Nova Cygni 1975 is a rare nova on a magnetic white dwarf (WD). Later it was found to be an asynchronous polar, now called V1500 Cyg. Our multisite photometric campaign occurring 40 years post eruption covered 26-nights (2015-2017). The reflection effect from the heated donor has decreased, but still dominates the optical radiation with an amplitude $\sim1^{m}.5$. The $0^{m}.3$ residual reveals cyclotron emission and ellipsoidal variations. Mean brightness modulation from night-to-night is used to measure the 9.6-d spin-orbit beat period that is due to changing accretion geometry including magnetic pole-switching of the flow. By subtracting the orbital and beat frequencies, spin-phase dependent light curves are obtained. The amplitude and profile of the WD spin light curves track the cyclotron emitting accretion regions on the WD and they vary systematically with beat phase. A weak intermittent signal at 0.137613-d is likely the spin period, which is 1.73(1) min shorter than the orbital period. The O-C diagram of light curve maxima displays phase jumps every one-half beat period, a characteristic of asynchronous polars. The first jump we interpret as pole switching between regions separated by $180^{\circ}$. Then the spot drifts during $\sim$ 0.1 beat phase before undergoing a second phase jump between spots separated by less than $180^{\circ}$. We trace the cooling of the still hot WD as revealed by the irradiated companion. The post nova evolution and spin-orbit asynchronism of V1500 Cyg continues to be a powerful laboratory for accretion flows onto magnetic white dwarfs.	
	
\end{abstract}




\section{Introduction}
Polars are interacting binaries consisting of a magnetic white dwarf (WD) accreting matter along field lines from a low mass main sequence secondary. The strong magnetic field of the WD prevents the formation of an accretion disk. The vast majority of polars are also synchronized, such that the 
spin of the WD becomes locked with the binary orbit due to the interaction of the magnetic field with the donor star, see Frank, King, and Raine (1992). 

Nova Cygni 1975 reached 2nd magnitude and thus was one of the brightest novae of the 20th century \citep[]{Honda1975, Warner2006}.
The nova occurred on the WD in the binary V1500 Cyg, later found to an asynchronous polar \citep[]{Schmidt1995}. The three other confirmed asynchronous polars are BY Cam, V1432 Aql, and CD Ind.
Of particular importance, V1500 Cyg is the site of a rare magnetic nova, along with GK Per and DQ Her. In the case of V1500 Cyg, it is assumed that the expanding nova shell slowed the orbit of the companion of a previously synchronized polar \citep[]{Stockman}. A strong irradiation of the secondary by the hot WD dominated the total radiation of this system for decades following the eruption. 

While polarization was observed to track the WD spin, optical photometry gave information on two periods: 1) the orbital modulation caused by heating of the secondary component by irradiation from the WD \citep[]{Schmidt1995} that is still very hot, 40 years after the 1975 explosion and 2) the spin-orbital (synodic) beat period \citep[]{PavPelt1991}. The orbital period has been detected every year of observation and its amplitude has varied from year to year \citep[]{PavShu1, PavShu2}. The beat period has systematically increased with time. Using the spin-orbit beat period relation, orbital and beat periods were used for $\it{calculation}$ of the current value of the WD spin period \citep[]{PavShu1} rather than direct measurements of the WD spin period which is possible only while the accretion flow is stable, between pole switchings, and is best done with polarimetric or X-ray observations.  

Spin-orbit synchronization times of a hundreds to a few thousand years are expected from theory, see e.g. \citet[]{Andronov1987}.  \citet[]{Campbell and Schwope1999} show that a range of synchronization times are possible with the fastest being about 50 years and requiring an unusually low-mass primary of 0.5 M$_\odot$. Such a low mass WD is unlikely given that evidence from the nova ejecta \citep[]{Lance1988} indicate a high mass WD. Photometry over the intervening years indicated that the WD was undergoing fast and nonlinear synchronization from measurements by \citet[]{Katz1991} with a rate of  $\dot{P} = 1.8\times10^{-6}$ for 1977-1979, $2.7\times10^{-8}$ for 1979-1987, and $2.4\times10^{-8}$ for  2000-2003 \citep[]{PavShu1}. \citet[]{Schmidt1995} obtained $\dot{P} = 3.86\times10^{-8}$ for 1987-1992 from polarimetry. The time of synchronization was estimated independently by \citet[]{Schmidt1995} as  $\sim170$ years and by \citet[]{PavPelt1991} as 150-290 years. Recently \citet[]{Harrison and Campbell2016}  suggest that  V1500 Cyg is already synchronized. Armed with extensive beat-phased resolved photometry we find that to the contrary, V1500 Cyg is still an asynchronous polar.

This work aims to unravel the combination of several sources contributing to the optical radiation of V1500 Cyg $\sim40$ years after the 1975 nova explosion and to establish the current status of V1500 Cyg in terms of synchronization and accretion geometry. First, one would expect a reduction of the reflection effect from cooling of the hot WD. Second, with reduced heating comes the appearance of significant, and recently detected, optical cyclotron radiation from accretion columns  \citep[]{Harrison and Campbell2018}. Third, ellipsoidal variations due to the aspherical Roche-lobe filling companion may be observed. In particular, the test for asynchronism is accomplished by examination of the variable orientation of the WD magnetic field with respect to the secondary component. In other words, asynchronous polars display light curves which vary according to the beat phase between the binary orbit and the WD spin and is confirmed by the detection of these periods.

\section{Observations}

CCD-photometry of V1500 Cyg was obtained 
during 25 nights within a 54-day interval in 2015, with a follow-up observation in 2017. Observations were carried out using five telescopes located  at four observatories. See the journal of observations given in Table~\ref{tab:t1_table}. Exposure times varied from 10-s with no significant dead-time between exposures using the Otto Struve 2.1 m telescope at McDonald Observatory to 60-180 s at the Crimean Astrophysical Observatory (CrAO) 1.25 m and Shajn 2.6 m telescopes as well as the Terskol 60 cm telescope and Sternberg Astronomical Institute 1.25 m. All of the images were unfiltered except for those obtained at McDonald Observatory, where a broad-band optical (BVR) filter was used, and on one night in 2017 where a B filter was used with the CrAO 2.6-m telescope. Science frames were dark subtracted and flat-fielded in the usual manner. The comparison star USNO B1 1381-0460911 \citep{Monet2003} which has B=19.05(5) and  V= 17.83(5) was used for absolute calibration while the stars C2 and C3 \citep{Kaluzhny1987} were used for differential photometry in 2015 and for B photometry in 2017. During the time of these observations the brightness of V1500 Cyg varied between $V = 19^{m}$ and $V = 21^{m}$. Photometric uncertainties varied depended on the size of telescope, the brightness level, and weather conditions. The maximum uncertainty was $\sim 0^{m}.1$. All of the data were converted to Heliocentric Julian Day (HJD) for further analysis.

\begin{figure}
\centering
	\includegraphics[width=3.0in,height=100mm]{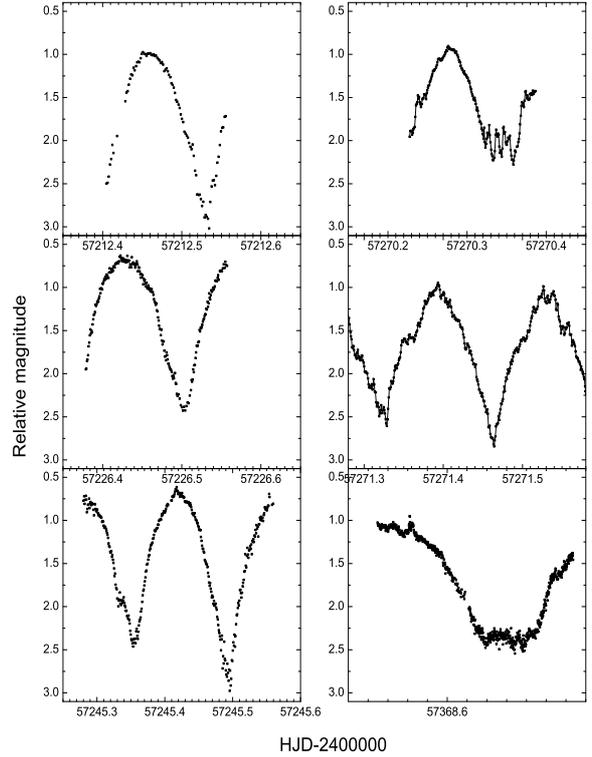}
	\caption{Examples of nightly light curves. All the data are presented in the same scale with the 
	exception of the highly time resolved data for HJD 2457368 showing the fine structure of the light curve. Light curves show significant night-to-night variability, typical of asynchronous polars.}
	\label{Fig1}
\end{figure}

\begin{table}
	\centering
	\caption{Journal of observations.}
	\label{tab:t1_table}
	\begin{tabular}{lccr} 
		\hline
		HJD 2457000+ & Observatory/ & CCD & N\\
		   (start - end) &telescope &  \\
		\hline
		211.407 - 211.564 & CrAO/2.6m & APOGEE E47&105\\
		212.405 - 212.555 & CrAO/2.6m & APOGEE E47 &94\\
		221.420 - 221.560 & CrAO/2.6m & APOGEE E47 &95\\
	    223.263 - 223.546 & CrAO/2.6m & APOGEE E47 &247\\
		225.380 - 225.553 & SAI/1.25m & VersArray-1300s & 186\\
		226.379 - 226.557 & SAI/1.25m & VersArray-1300s & 233\\
		245.280 - 245.559 & CrAO/2.6m & APOGEE E47 & 341\\
		270.228 - 270.386 & CrAO/2.6m & APOGEE E47 & 199\\
		271.264 - 271.584 & CrAO/2.6m & APOGEE E47 & 392\\
		272.225 - 272.479 & CrAO/2.6m & APOGEE E47 & 300\\
		274.221 - 274.378 & Terskol/60cm & SBIG STL 1001 & 138\\
		278.264 - 278.421 & Terskol/60cm & SBIG STL 1001 & 100\\
		281.266 - 281.493 & Terskol/60cm & SBIG STL 1001 & 135\\
		284.286 - 284.475 & Terskol/60cm & SBIG STL 1001 & 124\\
		295.302 - 295.437 & Terskol/60cm & SBIG STL 1001 & 87\\
		300.241 - 300.377 & CrAO/1.25m & ProLine PL23042 & 65\\
		301.231 - 301.419 & CrAO/1.25m & ProLine PL23042 & 69\\
		311.191 - 311.350 & CrAO/1.25m & ProLine PL23042 & 74\\
		326.188 - 326.374 & CrAO/1.25m & ProLine PL23042 & 73\\
		331.176 - 331.324 & CrAO/1.25m & ProLine PL23042 & 70\\
		332.196 - 332.348 & CrAO/1.25m & ProLine PL23042 & 70\\
		341.131 - 341.208 & SAI/1.25m & VersArray-1300s & 56\\
		344.133 - 344.344 & SAI/1.25m & VersArray-1300s & 146\\
		344.161 - 344.316 & CrAO/1.25m & ProLine PL23042 & 71\\
		367.559 - 367.637 & McDonald/2.1m & ProEM & 659\\
		368.565 - 368.664 & McDonald/2.1m & ProEM & 804\\
     	1013.389 - 1013.594 & CrAO/2.6m & APOGEE E47&127\\		
		\hline
	\end{tabular}
\end{table}











\section{Results}


Examples of nightly light curves are presented in  Fig.~\ref{Fig1}. The light curves commonly display smooth roundish maxima and sharp minima. Some light curves have step-like or teeth-like variations which distort the smooth intensity modulation. A majority of the light curves have an amplitude of about 1.5 mag. However on some occasions, the nightly amplitude was as small as 1 mag., see for example the light curve on HJD 2457270, top right panel of Fig.~\ref{Fig1}. Nightly light curves also display mean brightness level variations. V1500 Cyg demonstrates complex brightness variability where the dominating signal is the orbital modulation caused by the strongly irradiated secondary, hereafter described as a  ``reflection effect". Night-to-night variability is, in part, attributed to periodic changes in our view of the accretion flow geometry which depends on the spin-orbit beat phase.

\subsection{Beat period}

\begin{figure}
        \includegraphics[width=3.25in,height=40mm]{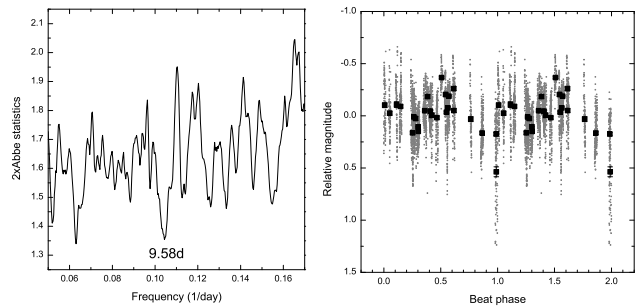}
	\caption{Left: periodogram for all of the data after orbital modulation is subtracted. Right: residuals (gray points) folded on the 9.58-d period. The mean per night values are marked by black squares. The zero phase corresponds to the start of observations (HJD =2457300.24118). For clarity the data are reproduced twice.}
	\label{Fig2}
\end{figure}

In order to investigate possible spin-orbital beat modulation we subtracted the orbital variability with a 0.13962-d period from the observed light curve and constructed a periodogram of the residual light curve using the ISDA package \citep[]{Pelt}.  The periodogram in the range of the expected beat period is shown as Fig.~\ref{Fig2}. The most significant feature is the minimum at a 9.58(3)-d period that is just a few percent longer than the predicted value of 9.1 - 9.3 d according to the ephemeris of \citet[]{Schmidt1995}, using polarimetry to track the spin of the WD and photometry for the orbital variability. The mean light curve has an amplitude of about 0.5 mag. So, the appearance of rapidly evolving light curves along with the determination of a putative beat period is strong evidence of spin-orbital asynchronism, still in 2015. 

\subsection{Orbital period: the reflection and possible ellipsoidal effects}

\begin{figure}
	 \includegraphics[width=3.25in,height=75mm]{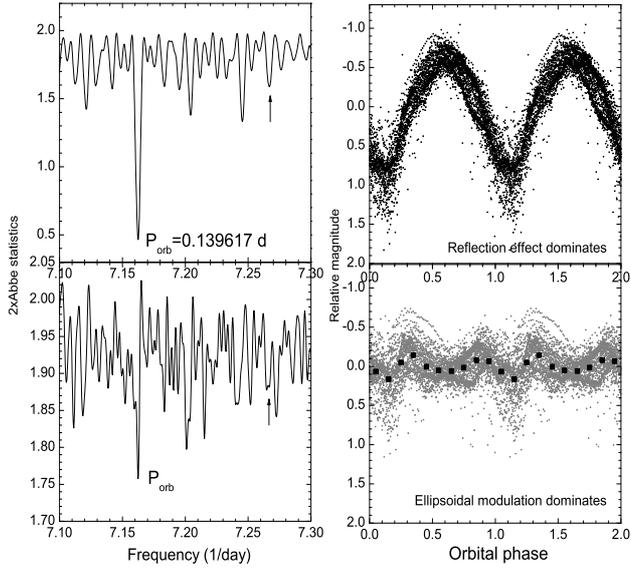}
	\caption{Top left: Periodogram for all of the data of V1500 Cyg after the beat modulation is subtracted reveals the strong signal at the orbital period. Top right: Corresponding data folded on the orbital period is consistent with a strong reflection effect. Bottom left: Periodogram for residuals after both beat modulation and orbital modulation are subtracted. Arrows in left panels point to the feature consistent with the predicted spin period by calculation. Bottom right: Corresponding data folded on the orbital period, now showing the combined cyclotron modulation with ellipsoidal variation from the aspherical companion. For clarity the data are reproduced twice in the right panels. Zero phase corresponds to the start of observations (HJD = 2457300.24118).}
	\label{Fig3}
\end{figure}

As we noticed that the dominant modulation is still caused by a strong reflection component, see top panels of Fig.~\ref{Fig3}, our data reveled a period of 0.139617(6)-d that within errors coincides with the orbital period measurement of 0.139613-d \citep[]{Semeniuk1995}. In order to extract weaker signals in the light curve we construct a periodogram of the residuals of all data after both the beat and orbital modulation (which we fit using a one-humped wave) were removed, see bottom panels of Fig.~\ref{Fig3}. The result reveled some excess at the orbital period in the form of a two-humped curve with near-equal maxima (amplitude of 0.35 mag). We suggest this is a light curve produced by the ellipsoidal shape of the secondary component in a field of high-mass white dwarf. However the true ellipsoidal profile will show a small difference from the one obtained due to some difference of a best-fitted curve from a properly modeled curve at a precise orbital inclination. Likely, the observed profile of the orbital modulation results from a superposition of reflection and ellipsoidal effects.

\subsection{Spin period}

\begin{figure}
        \includegraphics[width=3.25in,height=75mm]{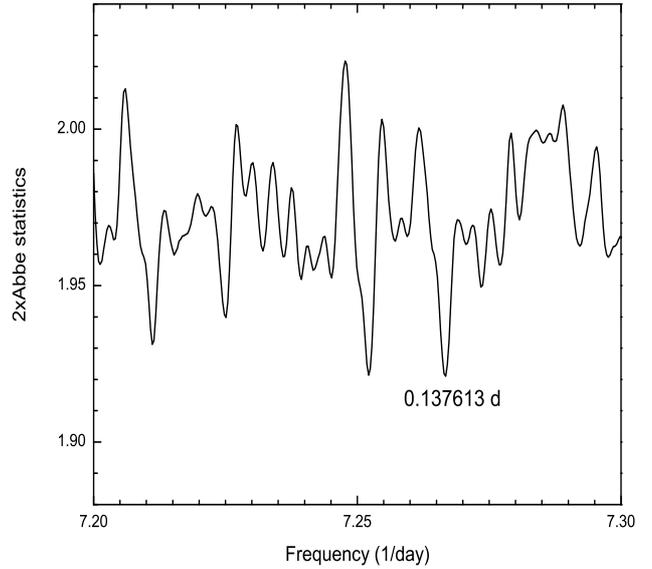}
	\caption{Periodogram for beat and orbit subtracted data in the region of the predicted spin period. This feature (recall arrows in Fig. 3) remains as other signals have weakened. 
This dilution of the spin signal is due to pole switching of the magnetically channeled accretion flow.}
	\label{Fig4}
\end{figure}

Inspection of the periodogram of the beat and orbital reflection subtracted light curve, see Fig.~\ref{Fig4}, reveals no prominent signal at the WD spin period, see arrows in left panels. Rather a residual signal is seen at the orbital period of 0.139617(6)-d. This signal is interpreted as the ellipsoidal modulation, an orbital variation due to the changing view of the aspherical Roche-lobe filling companion, as discussed in the previous subsection. After the ellipsoidal modulation is subtracted,  a weak WD spin signal at 0.137613(7)-d remains (see Fig.~\ref{Fig4}). As expected, the spin signal is strongly diluted by changes in the accretion flow geometry likely including pole switching. The marginally detected WD spin period is consistent with the calculated spin period Pspin according to equation~(\ref{eq:beat}), namely,

\begin{equation}
P_{beat}^{-1} = P_{spin}^{-1} - P_{orb}^{-1},
\label{eq:beat}
\end{equation}

where the observed orbital period is $P_{orb}$ = 0.139617-d and beat period is $P_{beat}$ = 9.58-d.
The dilemma in measuring the spin period directly is that one must either do so in less than one-half of a beat cycle or face significant signal dilution due to accretion pole switching. The WD spin period predicted using equation~(\ref{eq:beat}) is $P_{spin}$ = 0.137611-d.

It is not so obvious that we must detect the signal at the spin period. Its detectability probably depends on the accretion geometry in the V1500 Cyg system and uniformity of the data distribution. If the accretion stream switches from one magnetic pole to the other one and back during the beat cycle and if data are distributed uniformly in all beat phases, we would expect to see no spin period. On the other hand, in the case of no pole switching the spin signal must be strong enough to be detected. As expected, the cooling of the WD has resulted in the reduction of the still dominant reflection effect and the appearance of cyclotron emission producing a photometric signal modulated at the spin period. We therefore identify the 0.137613-d period as the WD spin period. Of particular significance, beat-phased resolved light curves indicate periodic changes in accretion geometry. We therefore conclude that while closer to being synchronous, the white dwarf in V1500 Cyg is still not synchronized and in 2015 its spin period was 173(1)-s or 0.88$\%$ shorter than its orbital period. 

\subsection{Accretion geometry}

\begin{figure}
	 \includegraphics[width=3.3in,height=175mm]{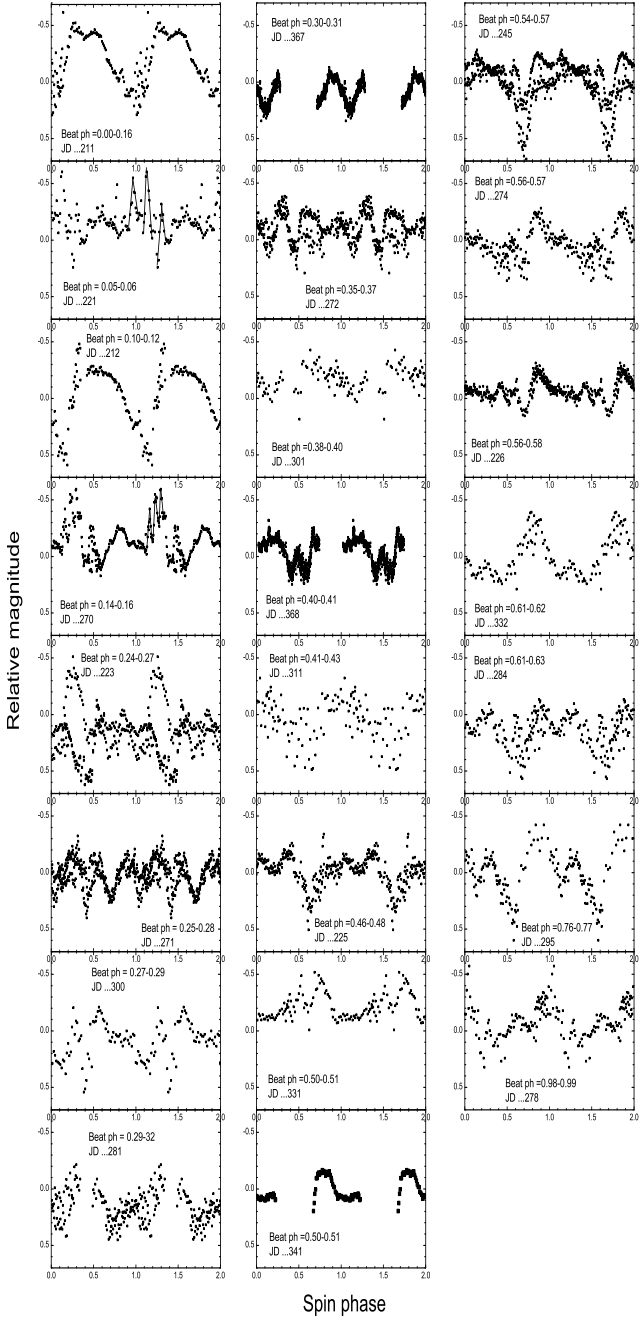}
	\caption{Photometric data folded on the adopted spin period of 0.137613 d. The beat and orbital period modulation were removed. The zero phase corresponds to the start of observations (HJD = 2457300.24118). For clarity the data are reproduced twice.}
	\label{Fig5}
\end{figure}

The residuals (after the beat and orbital modulations were subtracted) are folded on the suggested WD spin period of 0.137613-d and are presented in Fig.~\ref{Fig5} for each night. The corresponding phases of the 9.58-d beat period are indicated. It is seen that the residual light curves still have a variable amplitude and profile. Several morphological structures are observed, including one-, two-, or even three-humped structures. The shape of light curve even changes from cycle-to-cycle on some occasions (JD ...223).  Sometimes, a high-amplitude QPO-like structure is seen (JD ...221, JD ...270) superposed on some humps.  The phases of highest maxima and deepest minima also change. Such a diversity of profiles could be explained by a complex structure of the WD magnetic field. Such a complex accretion geometry as has been proposed for the asynchronous polar BY Cam \citep[]{Mason1998}.

\begin{figure}
 \includegraphics[width=3.25in,height=75mm]{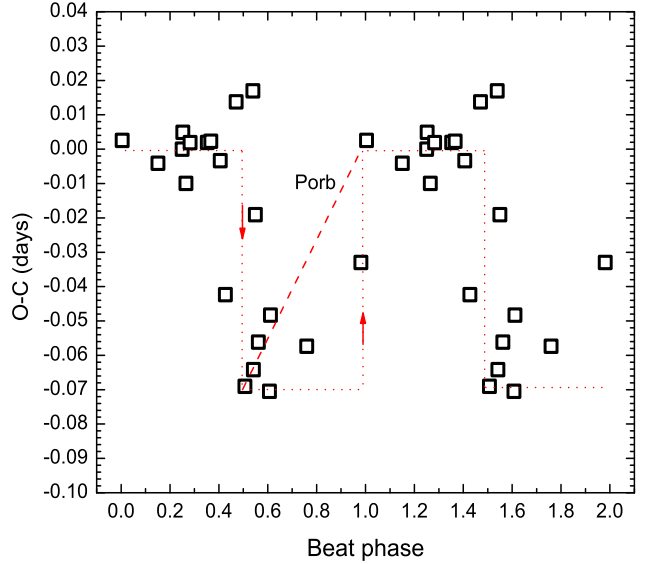}
\caption{The O-C diagram for WD spin maxima versus beat phase. The horizontal dotted red line is expected for an accretion spot that does not drift in position and thus varies at the P$_{spin}$ period of the WD. The vertical dotted lines correspond to expected jumps in O-C  at beat phases 0.0 and 0.5 that means  switching of accretion spot between poles separated by $180^{\circ}$. The O-C values are consistent with this model and also display a drifting behavior between beat phases 0.5 and 1.0.}
\label{Fig6}
\end{figure}

At first glance, changes in the nightly light curves appear chaotic. A more attentive procedure was performed in order to confirm or refute this suggestion. We determined times of the highest maxima (see Table~\ref{tab:t2_table}) assuming that they originate from the more intense accretion spot on the WD and calculated a O-C diagram using the ephemeris: 
$$HJD_{max} = 2457211.40731 + 0.137613 E.$$
All available maxima are plotted against corresponding phases of the beat period and are displayed in Fig.~\ref{Fig6}. This plot clearly shows two clusters of points in O-C space. The first cluster occupies half of the beat period  (beat phases $0.0 - 0.5$) and suggests that the observed period of the brightest accretion spot during that time is near constant and is equal to the adopted spin period. The position of the rest of the O-C values situated at beat phases $0.5 - 1.0$ imply that the accretion spot period is longer than the WD spin period, but shorter than orbital period, during that time.  There are two jumps occurring at beat phases 0.0 and 0.5 suggesting that the accretion stream switches between accretion spots at those phases. Further, during beat phases $\sim$ 0.5-0.62 the accretion spot drifts in the opposite direction to the WD rotation with a period close to the orbital one. At later phases it probably remains relatively constant and close to the spin period. Finally, at beat phase 1.0 the accretion stream switches to the second  region separated from the first accretion spot by less than $180^{\circ}$. Unfortunately, a lack of data does not allow complete coverage between beat phases 0.8 and 1.0. 

\begin{table}
	\centering
	\caption{Timings of spin maxima for residuals after the beat and orbital modulations were removed.}
	\centerline{HJD}
	\label{tab:t2_table}
	\begin{tabular}{lccr} 
		\hline
		2400000+ & 2400000+ & 2400000+ & 2400000+\\
		\hline
		57211.448 & 57223.280 & 57225.495 & 57245.315\\
		57245.416 & 57270.339 & 57271.311 & 57271.434\\
		57272.272 & 57272.410 & 57274.278 & 57278.292\\ 
		57284.332 & 57295.332 & 57300.345 & 57311.310\\
		57331.237 & 57332.199 & 57341.150 & 57368.596\\

		\hline
	\end{tabular}
\end{table}

\section{Discussion}

\begin{figure}
 \includegraphics[width=3.25in,height=75mm]{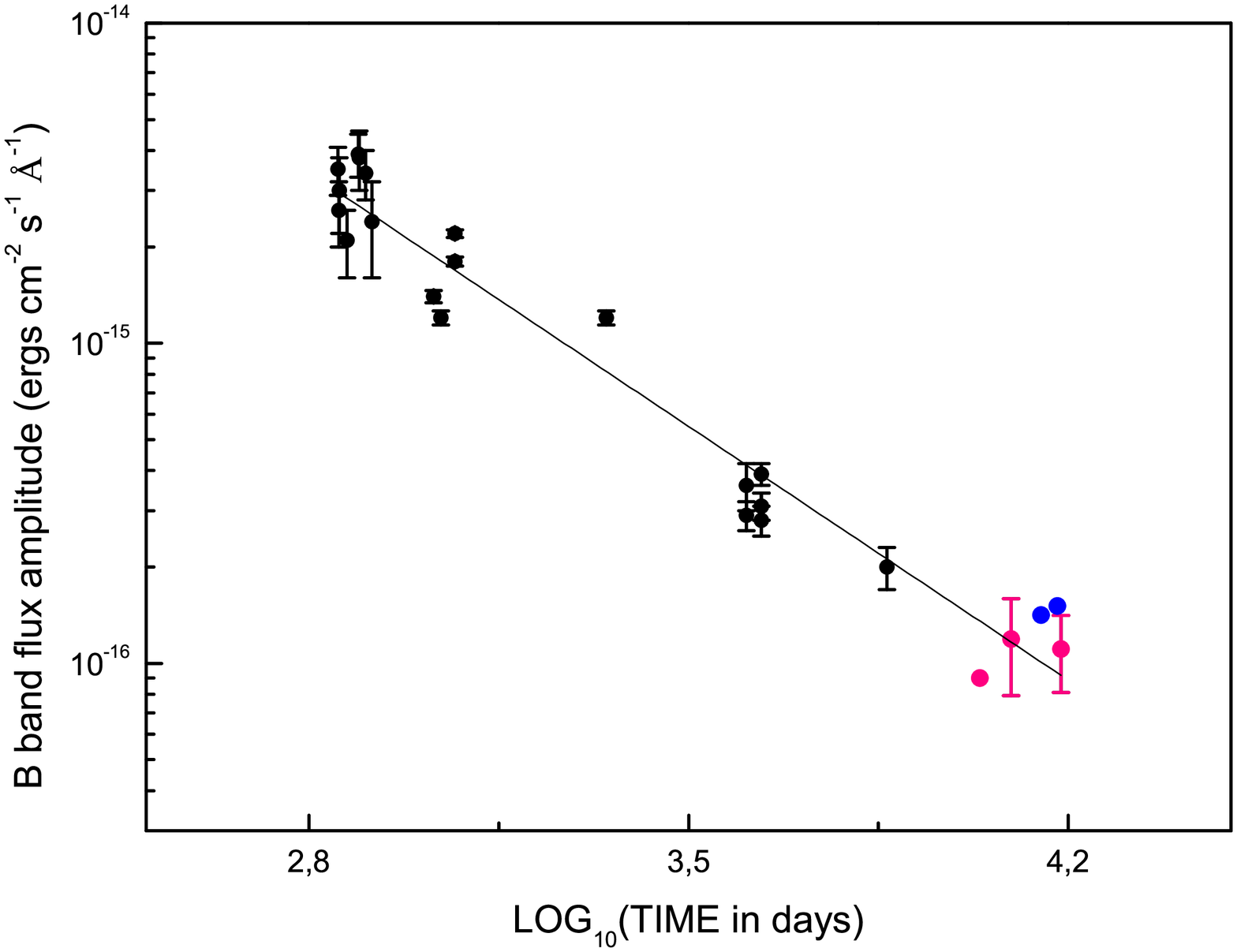}
\caption{Cooling of the white dwarf. 
The log of the amplitude, A, of the reflection modulation is shown as a function of the 
log of the time T since the nova eruption in 1975. Data is collected from the similar plot of 
\citet[]{Somers1999}, see note in text,  (black points) and more recent measurements of \citet[]{Litvinchova2011} and our new B-band light curve (pink points) and those of \citet[]{Harrison and Campbell2016} (blue points). Scatter is indicative of amplitude variation as a function of beat phase. The best-fit matches the theoretical prediction for WD cooling of \citet[]{Prialnik1986}}
\label{Fig7}
\end{figure}

In order to further investigate the cooling time-scale for the WD in V1500 Cyg, we gathered all
of the data on the amplitude of B-band photometry to follow the decrease in emission from the heated companion in the same manner as \citet[]{Somers1999}. We included data from their Table 1 along with data from \citet[]{Litvinchova2011} and \citet[]{Harrison and Campbell2016} and plotted them in Fig.~\ref{Fig7}. As it turns out, the new data lie on the extension of the previous relation 
confirming that the orbital amplitude decrease is indeed in accordance with
theoretical predictions for post-nova cooling. 
The best-fit power low index is $\eta$ = 1.13, which is remarkably consistent with the 
purely theoretical prediction of \citet[]{Prialnik1986}. Specifically, that the WD post-nova cooling should follow such a power-law with $\eta$ = 1.14, see the discussion in \citet[]{Somers1999}. 
The scattering around the best-fit line is caused by both photometric accuracy
and on the variable amplitude observed at different beat phases, recall the variations seen 
in Fig. \ref{Fig5}.

There is some confusion in the paper of \citet[]{Somers1999} concerning the quantities $\eta$ and x. The discrepancy is between their formulae and what is stated in the text.  According to their paper the response of the secondary star to heating is x=0.75.
Equating their equations (2) to (3) we obtain  $\eta$ =-1.68, not 0.94 as quoted in the text.  The value 0.94 could come from x=1/0.75,  or if the index in equation (3) is $\eta$/x  instead of $\eta$ $\times$ x. Our $\eta$, plotted in Fig.7 is also not the actual $\eta$, but rather the slope in that relation. We re-calculated this slope for Somers and Naylor data and found that it is -1.16, not -1.26.  Indeed, all the data give a slope of -1.13 that is reasonably close to -1.16. Despite this discrepancy, the slope of the amplitude is correctly measured and our new points are in agreement with the total trend.

Finally we consider the new observations of V1500 Cyg in comparison another asynchronous polar. BY Cam was the first asynchronous polar where evidence for accretion pole switching was found by polarimetric \citep[]{Mason1989} and photometric \citep[]{Mason1996, Silber1997, Mason1998} studies. A multi-polar magnetic field model of an aligned dipole plus quadrupole field model was suggested by \citet[]{Wu, Mason1998}. \citet[]{Pavlenko2006} found additional evidence of switching between a dipole spot and an equatorial one, with some drift of an accretion spot in the opposite direction to the WD spin. MHD calculations around the beat cycle were also performed demonstrating the main effects of pole switching in the case of a complex magnetic field structure \citep[]{Zhilkin}. 

The profiles of the 2015 spin light curves of V1500 Cyg are reminiscent of those of BY Cam in a high accretion state. Here for V1500 Cyg we find both the pole switching and a similar accretion spot drift as that observed in BY Cam. The major difference being, in V1500 Cyg, the domination of the optical light is from the hot WD irradiating the companion producing a strong reflection effect.  

\section{Conclusions}

We have carried out a photometric investigation of the asynchronous polar and post-nova V1500 Cyg  $\sim$40 years after its 1975 eruption. Our extensive dataset consists of photometric observations obtained during 25 nights within a 54-day interval in 2015 and during one night in 2017 at four observatories. Several key features in  the light curve argue for continued asynchronous spin-orbital motion of the the binary. These are the existence of nightly variations in the orbital light curve following the 9.6-d spin-orbital beat cycle, the enigmatic appearance of the 0.137613-d spin period which is only seen during some beat phases, and ultimately evidence of accretion stream flow switching between magnetic poles.


Earlier it was suggested \citep[]{Litvinchova2011} that the beat modulation in V1500 Cyg is caused by a periodic shading of the heated side of the secondary by an accretion ring around the WD. This explanation is in agreement with the interpretation of \citet[]{Schmidt1995}  of "relatively broad optically thick accretion columns with lumps of dense gas, and/or orbiting debris in a disk-like geometry". We can not distinguish between these possibilities, but we can say that such a structure would be a long-lived one, existing for the last 40 years. More high density observations covering the complete beat cycle are needed to monitor the path towards synchronism and to further investigate the beat phase dependent peculiarities in accretion geometry of V1500 Cyg.

An important question remains. Why is the WD in V1500 Cyg still so hot? After 40 years it remains hot enough for the irradiated secondary to dominate the optical flux, yet it CONTINUES TO FOLLOW THE THEORETICAL COOLING LAW. The nova eruption was among the most luminous ever recorded and apparently heated the WD to a higher temperature than is typical. Other magnetic CVs have also had luminous novae, namely GK Per and DQ Her, suggesting that magnetic novae may be more luminous as a sub-class. Further work is needed to properly address this question.

\section*{Acknowledgements}

P. Mason thanks the SDSS/FAST program for support. O. Antonyuk  acknowledges the Russian Science Foundation (grant N14-50-00043) and S. Shugarov thanks grants VEGA2/0008/17 and APVV 15-0458 for partial financial support.  





\bsp	
\label{lastpage}
\end{document}